\documentclass[]{interact}

\usepackage{epstopdf}
\usepackage{subfigure}

\usepackage[numbers,sort&compress,merge]{natbib}
\bibpunct[, ]{(}{)}{,}{n}{,}{,}
\renewcommand\bibfont{\fontsize{10}{12}\selectfont}
\renewcommand\citenumfont[1]{\textit{#1}}
\renewcommand\bibnumfmt[1]{(#1)}

\theoremstyle{plain}

\theoremstyle{definition}

\theoremstyle{remark}

\usepackage{psfrag}

\begin{document}

\articletype{ARTICLE TEMPLATE}

\title{Transient dynamics in cavity electromagnetically induced transparency with ion Coulomb crystals}

\author{
\name{Magnus Albert, Aur\'{e}lien Dantan and Michael Drewsen\thanks{CONTACT M. Drewsen. Email: drewsen@phys.au.dk}}
\affil{Department of Physics and Astronomy, Aarhus University, DK-8000 Aarhus C, Denmark}
}

\maketitle

\begin{abstract}
We experimentally investigate the transient dynamics of an optical cavity field interacting with large ion Coulomb crystals in a situation of electromagnetically induced transparency (EIT). EIT is achieved by injecting a probe field at the single photon level and a more intense control field with opposite circular polarization into the same mode of an optical cavity to couple Zeeman substates of a metastable level in $^{40}$Ca$^+$ ions. The EIT interaction dynamics are investigated both in the frequency-domain - by measuring the probe field steady state reflectivity spectrum - and in the time-domain - by measuring the progressive buildup of transparency. The experimental results are observed to be in excellent agreement with theoretical predictions taking into account the inhomogeneity of the control field in the interaction volume, and confirm the high degree of control on light-matter interaction that can be achieved with ion Coulomb crystals in optical cavities.
\end{abstract}


\section{Introduction}\label{sec:intro}

Electromagnetically induced transparency (EIT) is an interference phenomenon, which occurs when two electromagnetic fields resonantly excite two different transitions sharing a common state. EIT has been investigated in various coherent media to control the propagation of light~\cite{Harris1997,Lukin2001,Fleischhauer2005,Lvovsky2009,You2011,Kurizki2015}. Since its first observation in atomic vapors~\cite{Boller1991}, the EIT-induced modification of the absorption and dispersion of various atomic media has been extensively exploited, \textit{e.g.}, for light storage and retrieval~\cite{Lukin2001,Fleischhauer2005,Lvovsky2009,Simon2010}.

Enclosing the atomic medium in an optical cavity can substantially enhance the effects of the EIT interaction, which can be used, \textit{e.g.}, for spectroscopy purposes~\cite{Scully1992,Budker1999,Shahriar2007}, to improve the efficiency of light storage~\cite{Lukin2000,Dantan2004,Gorshkov2007,Boozer2007,Bimbard2014}, to realize low-light-level optical switching~\cite{Tanji-Suzuki2011}, to enhance optical nonlinearities~\cite{Imamoglu1997,Grangier1998,Gheri1999,Werner1999,Dantan2006b,Nikoghosyan2010} or for optomechanical cooling~\cite{Genes2011,Bienert2012,Kampschulte2014}. 

Cavity EIT has been observed with atomic beams~\cite{Muller1997}, in cold and room-temperature atomic vapors~\cite{Lukin1998,Wang2000,Hernandez2007,Wu2008,Zhang2008,Laupretre2011,Tanji-Suzuki2011}, cold Rydberg ensembles~\cite{Parigi2012,Ningyuan2016,Boddeda2016}, with ion Coulomb crystals~\cite{Albert2011}, and even with single or few atoms~\cite{Mucke2010,Kampschulte2010,Kampschulte2014}.
In almost all these experiments - Ref.~\cite{Laupretre2011} excepted - cavity EIT is observed via the transparency window appearing in the cavity steady state transmission or reflection spectrum. The narrowing of the cavity linewidth is related to the slowing down of the group velocity of the intracavity probe field in presence of the EIT medium. In a time-domain picture the width of the transparency window can be interpreted as the inverse of the effective cavity field lifetime~\cite{Laupretre2011} or the inverse of the time it takes for the electromagnetically induced transparency to build up inside the cavity~\cite{Dantan2006}.

In this work, we experimentally investigate transient cavity field dynamics in EIT with large ensembles of trapped, laser-cooled ions, forming so-called ion Coulomb crystals (ICCs)~\cite{Wineland1987,Diedrich1987,Drewsen1998}. Following the experiments of Ref.~\cite{Albert2011}, EIT is achieved by injecting a probe field at the single photon level as well as a more intense control field with opposite circular polarization into the same mode of an optical cavity to couple Zeeman substates of the D$_{3/2}$ metastable level in $^{40}$Ca$^+$ ions (see Fig.~\ref{fig:levels}). Applying this scheme to large ICCs containing several thousands of ions and strongly coupled to the cavity field~\cite{Herskind2009NatPhys}, changes from essentially full transmission to full absorption of the probe field was demonstrated in~\cite{Albert2011}. Here, the cavity EIT interaction dynamics are investigated both in the frequency-domain - by measuring the probe field steady state reflectivity spectrum - and in the time-domain - by measuring the progressive buildup of the transparency on two-photon resonance. The widths of the transparency windows appearing in the steady state spectrum and the EIT buildup rates are compared for various control field intensities and ICCs with different coherent couplings with the cavity. The results are also compared to analytical and numerical theoretical predictions~\cite{Dantan2012}, which take into account a specificity of this ``all-cavity" EIT scheme, namely the inhomogeneity of the control field intensity in the interaction volume. The results between the experimental data and the theoretical predictions are in excellent agreement, confirming that ICCs in optical cavity represent a very well-controlled system for cavity quantum electrodynamics investigations. They are in particular promising for the realization of photon memories~\cite{Zangenberg2012} or counters~\cite{Clausen2013} based on ICCs. Together with those of Ref.~\cite{Albert2011}, the results presented here represent an important extension of previous work focussing on transient EIT dynamics in single-pass experiments~\cite{Fry1993,Harris1995,Li1995,Chen1998,deEchaniz2001,Greentree2002}.

\section{Cavity EIT}\label{sec:theory}

\begin{figure}
\centering
\subfigure[ ]{
\resizebox*{5cm}{!}{\includegraphics{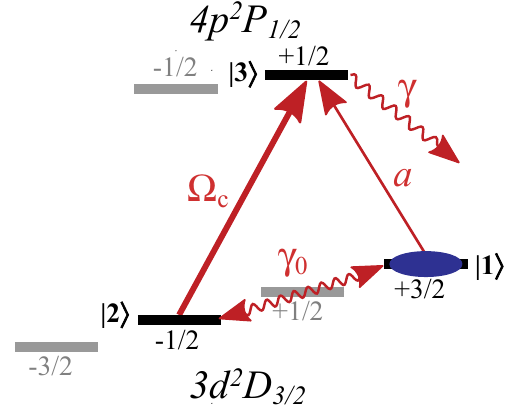}}}\hspace{4pt}
\subfigure[ ]{
\resizebox*{5.5cm}{!}{\includegraphics{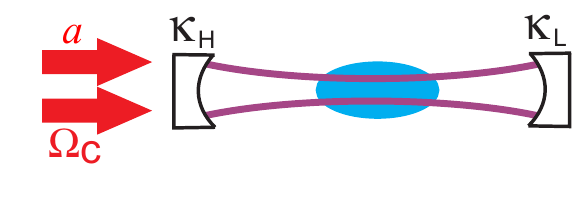}}}
\caption{Level scheme (a) and cavity configuration (b) considered.}\label{fig:levels}
\end{figure}

To set the basis for the discussion let us first consider an ensemble of three-level $\Lambda$ atoms with ground states $|1\rangle$ and $|2\rangle$ and excited state $|3\rangle$, corresponding to the Zeeman sublevels of interest of states in $^{40}$Ca$^+$ ions (Fig.~\ref{fig:levels}a). The ions interact with two field modes of a linear Farby-Perot cavity (Fig.~\ref{fig:levels}b): the probe field on the $|1\rangle\longrightarrow|3\rangle$ transition and the control field on the $|2\rangle\longrightarrow|3\rangle$. The maximum single-ion coupling rate with the probe and control fields at the center of the cavity are denoted by $g$ and $g_c$, respectively. We assume that the length of the ion ensemble is much smaller than the Rayleigh range of the cavity modes and hence neglect longitudinal variations of the cavity field modefunctions. We further assume that the thermal motion of the randomly distributed ions along the cavity axis results in averaged longitudinal couplings $\bar{g}=g/\sqrt{2}$ and $\bar{g}_c=g_c/\sqrt{2}$~\cite{Herskind2009NatPhys,Albert2012,Dantan2012}. For an ensemble whose transverse extension is larger than the cavity field waists, the coupling strengths of ions at a distance $r_j$ from the cavity axis have to weighted by the amplitude of the modefunctions of the probe and control fields, $\Psi_p(r_j)$ and $\Psi_c(r_j)$. We assume additionally that all ions are initially prepared in state $|1\rangle$ and that the control field, resonant with the $|2\rangle\longrightarrow|3\rangle$ transition, is injected into the cavity, whose frequency is resonant with the $|1\rangle\longrightarrow|3\rangle$, and that the steady state number of control field intracavity photons is much larger than the number of intracavity probe photons, which is of order unity. In this ``weak" probe regime, in which almost all the ions stay in state $|1\rangle$, the semiclassical equations describing the dynamics of the ions and the probe field in this regime are given by~\cite{Dantan2012}
\begin{align}
\dot{a} &= -(\kappa-i\Delta)a+i\bar{g}\sum_j\Psi_p(r_j)\sigma^{(j)}+\sqrt{2\kappa_H}a_{\textrm{in}}\\
\dot{\sigma}^{(j)} &= -(\gamma-i\delta)\sigma^{(j)}+i\bar{g}\Psi_p(r_j)a+i\bar{\Omega}_c\Psi_c(r_j)s^{(j)}\\
\dot{s}^{(j)} &=-(\gamma_0-i\Delta)s^{(j)}+i\bar{\Omega}_c^*\Psi_c(r_j)\sigma^{(j)}
\end{align}
with $a=\langle \hat{a}\rangle$, $\sigma^{(j)}=\langle\hat{\sigma}^{(j)}\rangle$ and $s^{(j)}=\langle\hat{s}^{(j)}\rangle$, where $\hat{a}$ is the annihilation operator in the rotating frame associated with probe field with frequency $\omega_p$, $\hat{\sigma}^{(j)}=|1\rangle\langle 3|$ is the dipole operator and $\hat{\sigma}^{(j)}=|1\rangle\langle 2|$ the ground state lowering operator associated with the $j$-th atom at position $r_j$. $\kappa$ is the cavity field total decay rate, while $\kappa_H$ is the cavity field decay rate through the incoupling mirror. $a_{\textrm{in}}$ is the incoming probe field amplitude (in $\sqrt{\textrm{Hz}}$ units). $\gamma$ is the optical dipole decay rate and $\gamma_0$ the ground state decoherence rate. $\Delta=\omega_p-\omega_{31}$ is the probe field one-photon detuning, $\omega_{31}$ being the frequency of the $|1\rangle\longrightarrow|3\rangle$ transition. As a consequence of our assumptions, $\Delta$ is also equal here to the cavity detuning and the two-photon detuning. $\bar{\Omega}_c=\bar{g}_c\langle a_c\rangle=\Omega_c/\sqrt{2}$ is the rescaled control field Rabi frequency. As in the experiments we take modefunctions corresponding to the fundamental mode of the cavity: $\Psi_{p,c}(r)=\exp(-r^2/w_{p,c}^2)$, where $w_{p}$ and $w_c$ are the probe and control field waists, respectively.

Solving these equations in steady state yields an intracavity probe field amplitude
\begin{equation}
a=\frac{\sqrt{2\kappa_H}a_{\textrm{in}}}{\kappa-i\Delta-i\chi_{\textrm{EIT}}}
\end{equation}
where the EIT susceptibility is given by
\begin{equation}
\chi_{\textrm{EIT}}=\sum_j\frac{i\bar{g}^2\Psi_p^2(r_j)}{\gamma-i\Delta+\frac{|\bar{\Omega}_c|^2}{\gamma_0-i\Delta}\Psi_c(r_j)^2}
\end{equation}
Assimilating the ion ensemble as a continuous medium with uniform density $\rho$ yields a susceptibility~\cite{Albert2011,Dantan2012}
\begin{equation}
\chi_{\textrm{EIT}}=\frac{ig_N^2}{\gamma-i\Delta}\frac{\ln(1+\Theta)}{\Theta}
\end{equation}
where
\begin{equation}
\Theta=\frac{\Omega_c^2/2}{(\gamma-i\Delta)(\gamma_0-i\Delta)}
\end{equation}
is an effective saturation parameter for the two-photon transition and $g_N=g\sqrt{N}$ is the \textit{collective coupling rate} with $N=\rho\frac{\pi w^2}{2}L$ the \textit{effective number of ions} introduced in~\cite{Herskind2009NatPhys,Albert2012}, $L$ being the half-length of the ion crystal and $w=w_p=w_c$ the cavity waist (assumed much smaller than the transverse extension of the crystal).

Using the relations $a_{\textrm{ref}}=\sqrt{2\kappa_H}a-a_{\textrm{in}}$ and $a_{\textrm{tr}}=\sqrt{2\kappa_L}a$, the probe field steady state transmittivity $\mathcal{T}=|a_{\textrm{tr}}/a_{\textrm{in}}|^2$ and reflectivity $\mathcal{R}=|a_{\textrm{ref}}/a_{\textrm{in}}|^2$ can be straightforwardly computed.

This ``all-cavity" situation can be compared to the canonical situation where all the atoms experience the same control field intensity. Taking $\Psi_c(r_j)=1$ yields an EIT susceptibility~\cite{Fleischhauer2005}
\begin{equation}
\chi_{\textrm{EIT}}^0=\frac{ig_N^2}{\gamma-i\Delta}\frac{1}{1+\Theta}
\end{equation}
In the regime in which the \textit{cooperativity} $C=g_N^2/(2\kappa\gamma)$ of the probe field with the ensemble is large, and for $\Omega_c\ll g_N$, the probe field transmission spectrum around two-photon resonance ($\Delta\simeq 0$) is well-approximated by a Lorentzian function of the detuning with a half-width at half-maximum (HWHM) given by
\begin{equation}
\gamma_{\textrm{EIT}}^0=\gamma_0+\frac{\Omega_c^2/2}{(1+2C)}
\end{equation}
which increases linearly with the control field intensity ($\varpropto\Omega_c^2$) and decreases with the effective number of atoms ($C\varpropto N$). The lineshapes obtained in the ``all-cavity" situation are markedly different from those of the canonical situation, although the same scalings can be observed, as has been discussed in~\cite{Albert2011,Dantan2012} and will be discussed further in Sec.~\ref{sec:results}.

Equivalently, one can look at the dynamics of the intracavity field buildup in EIT. In the canonical situation of an homogeneous control field and in the regime $C\gg 1$, $\Omega_c\ll g_N$ considered previously, it can be straightforwardly shown that the intracavity probe field and the collective optical dipole adiabatically follow the collective ground state coherence, which builds up exponentially with a rate $\gamma_{\textrm{EIT}}^0$, consistently with the frequency-domain picture discussed above. The dynamics in the ``all-cavity" situation are more complex, due to the inhomogeneity of the control field intensity, and need to be solved numerically, but show  approximately the same scalings with the control field intensity and the effective number of ions, given an appropriate rescaling of the control field Rabi frequency, as will be discussed in more detail later.

In the next sections we investigate both frequency- and time-domain pictures experimentally.

\section{Experimental setup and sequence}\label{sec:setup}

\begin{figure}
\centering
\includegraphics[width=0.32\textwidth, angle=270]{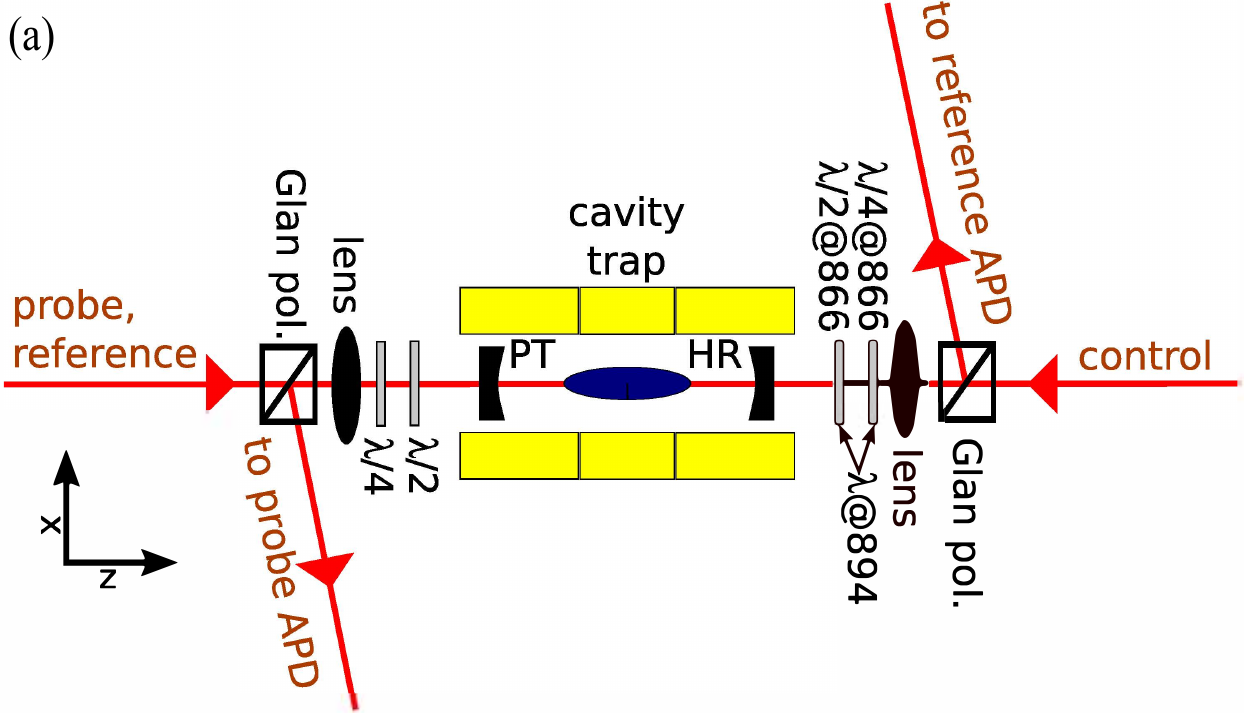}\\\vspace{5pt}
\includegraphics[width=0.6\textwidth]{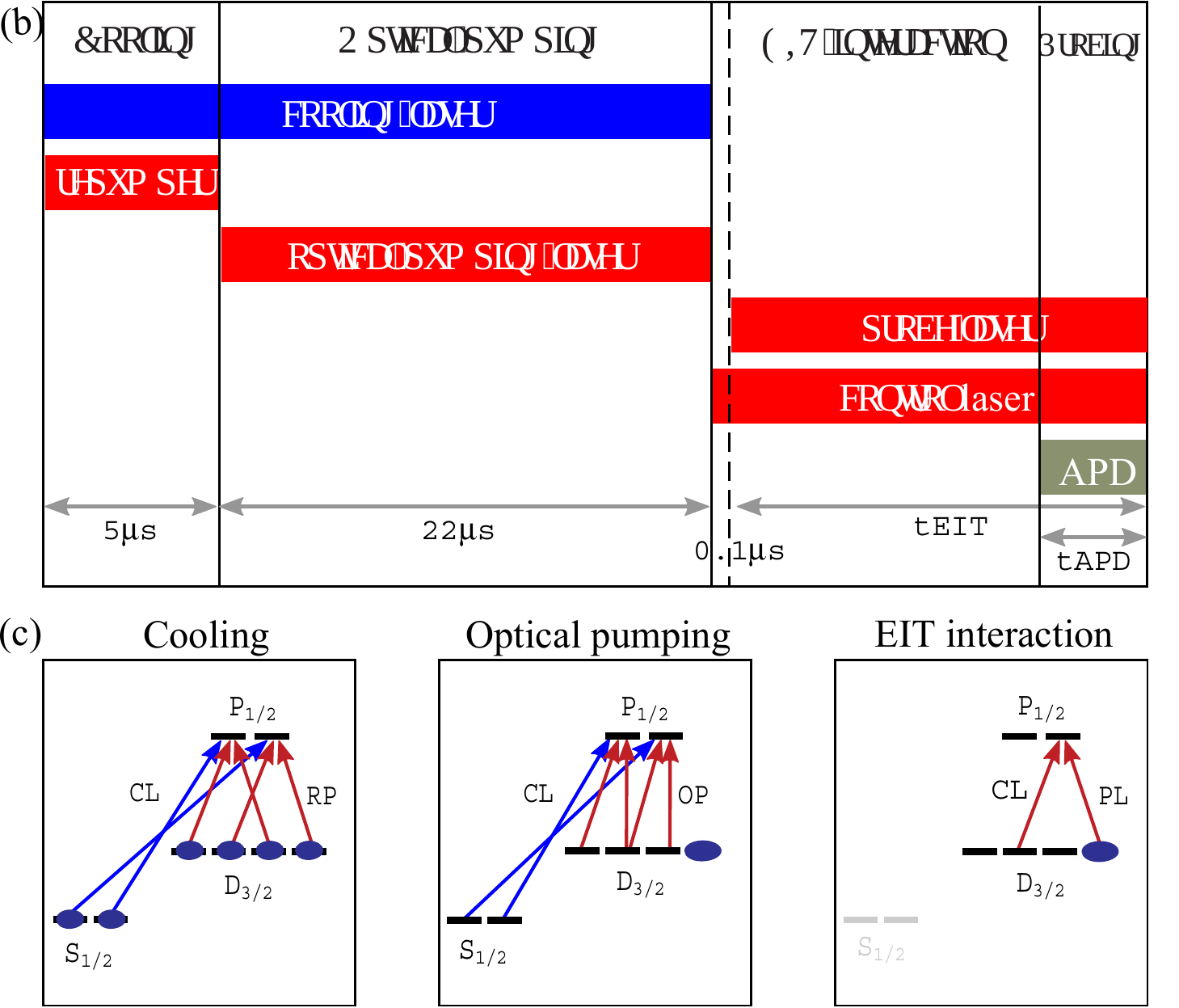}
\caption{(a) Schematic of the experimental setup. PT: high-transmission mirror, HR: low-transmission mirror, APD: avalanche photodiode. (b) Layout of the experimental sequence. (c) Relevant energy levels of $^{40}$Ca$^+$ and driven transitions involved during the different phases of the sequence. CL: cooling laser, RP: repumper laser, OP: optical pumping laser, CL: control field laser, PL: probe field laser.}\label{fig:sequence}
\end{figure}

A schematic of the experimental setup is shown in Fig.~\ref{fig:sequence}a. The cavity ion-trap used for these experiments has been described in detail in~\cite{Herskind2008,Herskind2009NatPhys,Herskind2009JPB,Albert2012}. In brief, large ICCs containing a few thousands of $^{40}$Ca$^+$ ions are loaded and laser cooled to temperatures of a few tens of millikelvins into a linear Paul trap in a quadrupole configuration. The crystals are typically prolate spheroids with $\sim$mm lengths along the longitudinal axis of the trap, diameters of a few hundreds of micrometers and uniform densities of a few $10^8$ cm$^{-3}$. A moderate finesse ($\sim 3000$ at 866 nm), 11.7 mm-long, close-to-confocal cavity is integrated into the trap. The cavity axis coincides with the longitudinal axis of the trap~\cite{Herskind2009JPB}, along which a $\sim 2.2$ G magnetic field is applied. The cavity mirrors have asymmetric intensity transmissions, $T_H=1500$ ppm and $T_L=4$ ppm, respectively. Taking into account an additional total roundtrip loss of $A\simeq 650$ ppm, the bare cavity field amplitude decay rate is $\kappa/(2\pi)=2.2$ MHz at a wavelength of 866 nm corresponding to the $3d ^2D_{3/2}$ to $4p ^2P_{1/2}$ transition in $^{40}$Ca$^+$. The waist (raidus) of the cavity at its center is 37 $\mu$m.

The experimental sequence to realize and investigate cavity EIT dynamics in the ion crystal follows that of Ref.~\cite{Albert2011} and is shown in Figs.~\ref{fig:sequence}b and c. The ions are first Doppler cooled and optically pumped into the $m_J=+3/2$ Zeeman substate of the $3d ^2D_{3/2}$ level by application of suitably detuned and polarized 397 nm and 866 nm light fields~\cite{Herskind2009NatPhys,Albert2011}. The weak probe field (single photon level) and the more intense control field, which are issued from the same laser, are both tuned close to resonance with the $3d ^2D_{3/2}$ to $4p ^2P_{1/2}$ transition, but with opposite circular polarizations to address the $m_J=+3/2$ and $m_J=-1/2$ Zeeman substates of the $3d ^2D_{3/2}$ level, respectively. The probe field is injected from the high-transmission mirror side and its reflection from the cavity is detected by an avalanche photodiode after spectral and spatial filtering. The intensity in the probe field is such that the mean number of steady state intracavity photons in the bare, resonant cavity is of the order of unity. The control field is injected from the low-transmission side and separated from the reflected probe field by a high-extinction Glan laser polarizer. An additional reference laser (wavelength 894 nm), which is far-off resonant from any transition in the atoms, is coupled into the cavity from the high-transmission mirror side and its transmission monitored to actively stabilize the cavity length. In the experiments reported here the cavity is kept on resonance with the $3d ^2D_{3/2}, m_J=+3/2$ to $4p ^2P_{1/2}, m_J=+1/2$ transition, while the probe field frequency is varied around resonance, thus allowing for probing the coupled modes of the ion crystal-cavity system.

As shown in Fig.~\ref{fig:sequence}b the control field, resonant with the $3d ^2D_{3/2}, m_J=-1/2$ to $4p ^2P_{1/2}, m_J=+/2$ transition, is injected 0.1 $\mu$s before the probe field to allow for its intracavity intensity to reach its steady state value (cavity intensity buildup/decay time constant $\sim$ 36 ns). The probe field is then injected into the cavity and both fields are applied for a duration $t_{\textrm{EIT}}$. At the end of this period the probe field reflectivity level is measured for $t_{\textrm{APD}}$ (1.4 $\mu$s or 0.5 $\mu$s, depending on the type of measurement).

\section{Results}\label{sec:results}

\subsection{Normal mode splitting}
\begin{figure}
\centering
\subfigure{
\resizebox*{6.5cm}{!}{\includegraphics{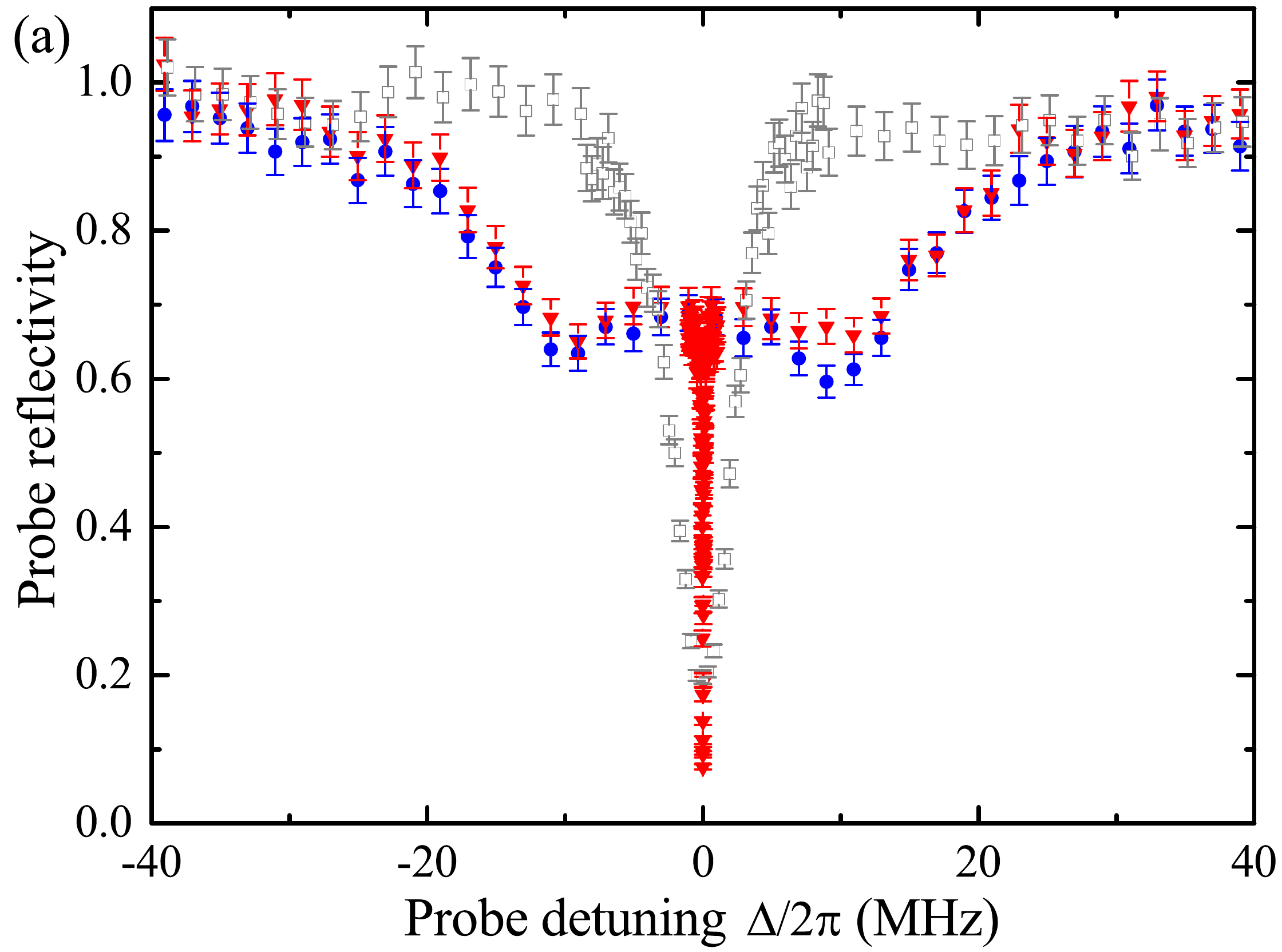}}}\hspace{5pt}
\subfigure{
\resizebox*{6.5cm}{!}{\includegraphics{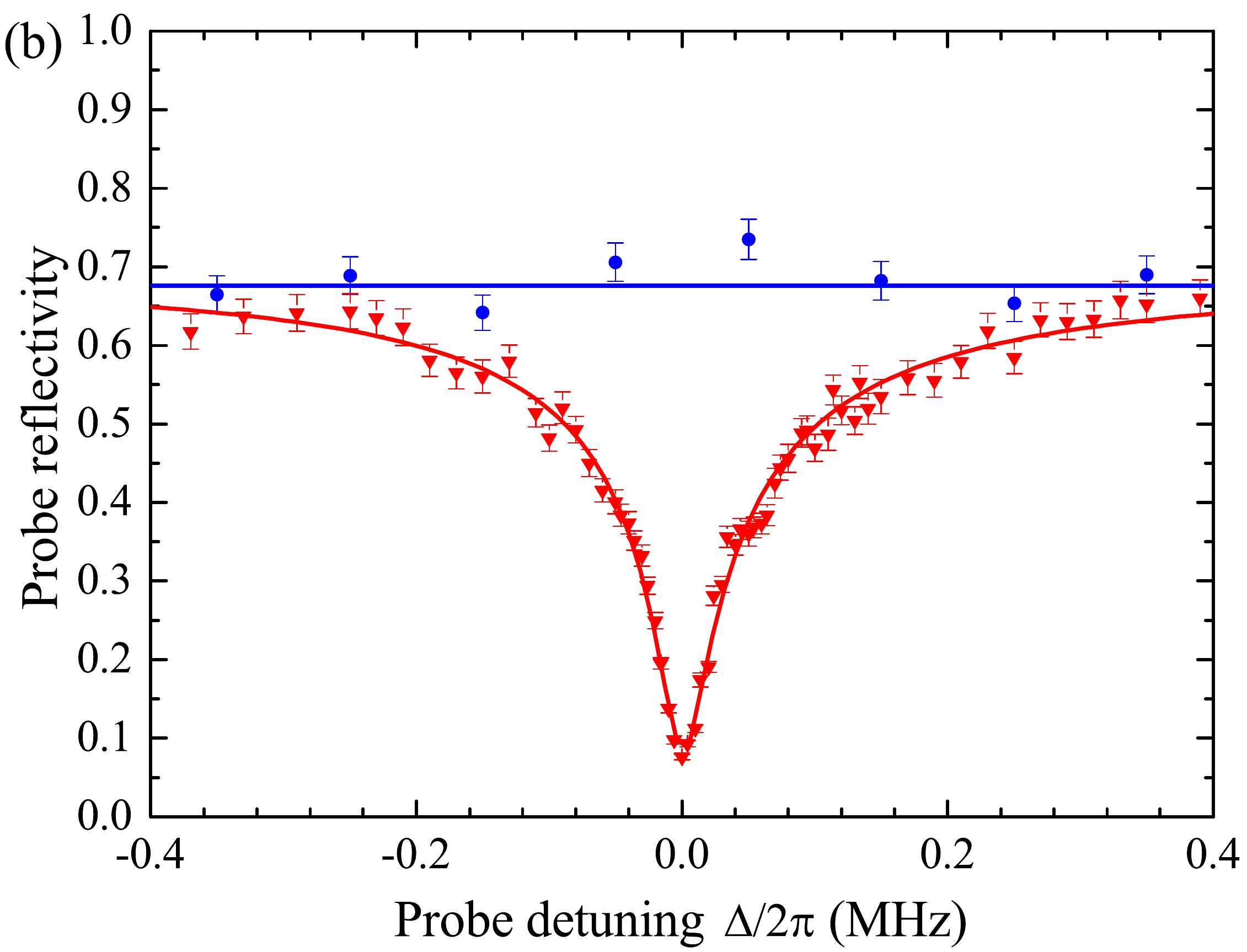}}}
\caption{(a) Probe reflectivity spectrum in absence (blue circles) and presence of a control field (red triangles) for a cavity containing an ICC (cooperativity $C\simeq 5.4$)). The grey squares shows the bare cavity spectrum for reference. (b) Enlarged view of the spectra around two-photon resonance. The solid lines show the results of theoretical fits.} \label{fig:rabi_EIT_spectra}
\end{figure}

\begin{figure}
\centering
\includegraphics[width=0.5\textwidth]{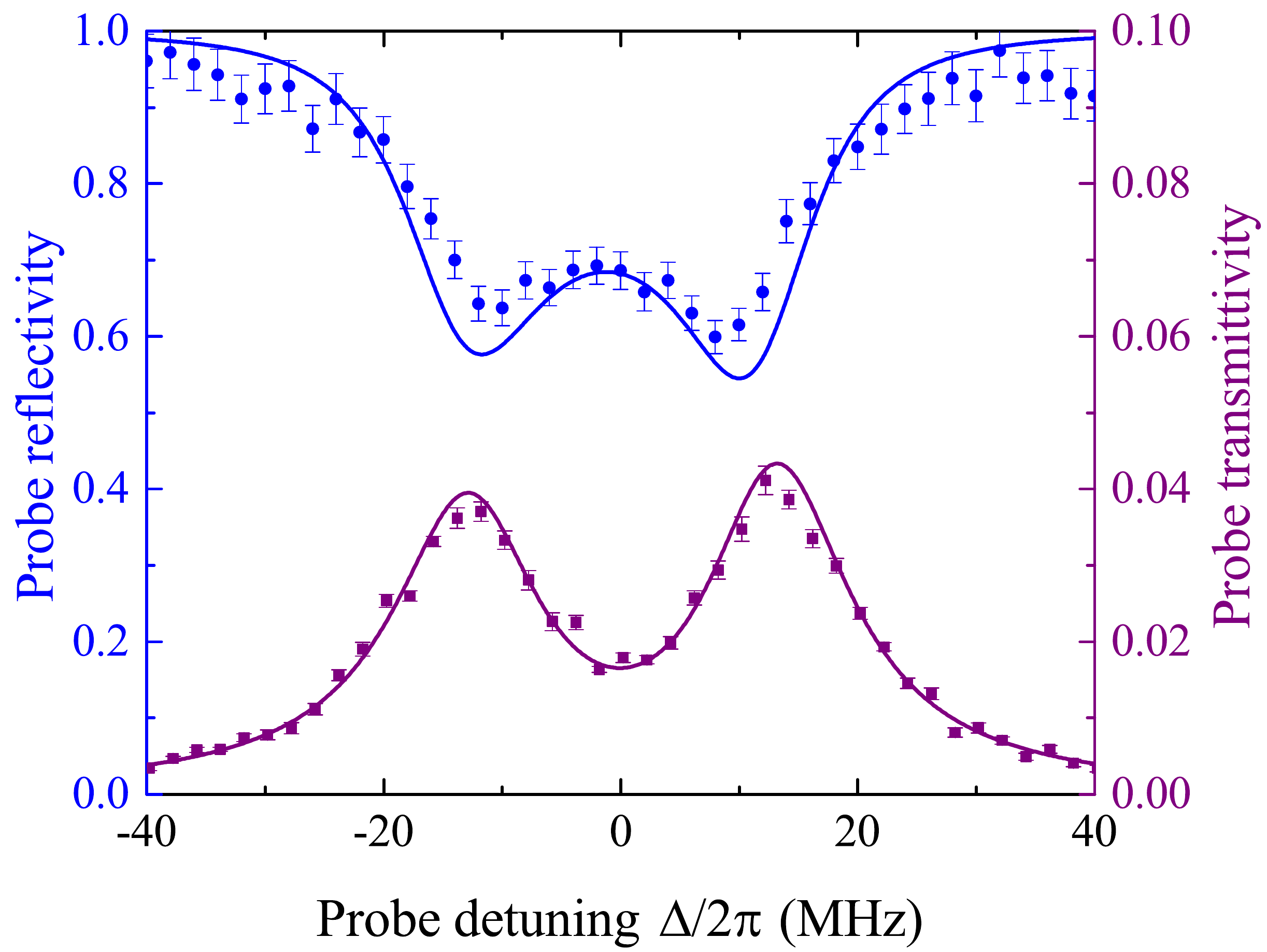}
\caption{Probe reflectivity and transmittivity spectra for a cavity containing the same ICC as in Fig.~\ref{fig:rabi_EIT_spectra}, in absence of control field. The lines show the results of theoretical fits. Notice the different scales for reflectivity and transmittivity.} \label{fig:rabi_ref_trans}
\end{figure}

Figure~\ref{fig:rabi_EIT_spectra} shows typical measured probe steady state reflectivity spectra, with and without ICC in the cavity and in presence or in absence of control field. For a bare cavity (grey squares) a Lorentzian spectrum is measured with a HWHM corresponding to cavity field decay rate of $\kappa=2\pi(2.2\pm 0.1)$ MHz. The blue circles correspond to the situation in which an ICC with length $1602\pm 2$ $\mu$m, diameter $282\pm2$ $\mu$m and density $(5.6\pm 0.1)\times 10^8$ cm$^{-3}$, is loaded into the cavity and only the probe field is applied. The dimensions are determined from fluorescence images of the crystal during laser cooling~\cite{Herskind2008} and the density from a careful calibration of the trapping potentials~\cite{Herskind2009JPB}. Such a crystal thus contains in total $\sim 37500$ ions out of which $\sim 680$ effectively interact with the cavity field~\cite{Herskind2009NatPhys,Albert2012}, yielding an expected collective coupling rate of $g_N^{\textrm{(th)}}=2\pi(13.8\pm0.1)$ MHz. Two dips are observed in the reflectivity spectrum -- a signature of the collective strong coupling regime between the ions and the probe field~\cite{Herskind2009NatPhys}, in which the collective coupling rate $g_N^{\textrm{(refl)}}=2\pi(13.9\pm0.3)$ MHz, obtained from a fit to the model (Fig.~\ref{fig:rabi_ref_trans}) supersedes both the spontaneous emission rate $\gamma$ and the cavity field decay rate $\kappa$. To perform this fit the cavity field decay rate $\kappa=2\pi(2.2\pm0.1)$ MHz and the effective spontaneous emission rate $\gamma=2\pi(12.6\pm 0.5)$ MHz were fixed to values independently determined by measurements of the ion-crystal broadened cavity linewidth spectrum~\cite{Herskind2009NatPhys,Albert2012}, which allow for evaluating the slight Doppler broadening of the probed transition due to the ions' thermal motion. 

To confirm the collective coupling rate value, the cavity probe spectrum were measured for the same crystal \textit{both} in reflection and transmission. The measurements of the transmittivity was performed as depicted in Fig.~\ref{fig:sequence}, but with the probe field being injected from the low-transmission mirror side instead of the control field, albeit with $\sigma_-$-circular polarization. Fits of the reflectivity and transmittivity spectra yield $g_N^{\textrm{(refl)}}=2\pi(13.9\pm0.3)$ MHz and $g_N^{\textrm{(tr)}}=2\pi(13.8\pm0.1)$ MHz, respectively. 

The red triangles show the probe reflectivity spectrum in presence of a control field with an intensity corresponding to $\sim500$ intracavity photons. In addition to the normal modes of the strongly coupled system described earlier, a narrow dip with HWHM $47.5\pm2.4$ kHz is observed around two-photon resonance, which defines a transparency window for the probe field. As discussed in~\cite{Albert2011} the atomic transparency at two-photon resonance is increased from 2\% in absence of control field to 84\% in its presence. A fit to the theoretical model described in Sec.~\ref{sec:theory} yields $g_N=2\pi(13.6\pm0.3)$ MHz and $\Omega_c=2\pi(4.1\pm0.1)$ MHz, in good agreement with the expected collective coupling rate and the control field Rabi frequency $\Omega_c^{\textrm{(th)}}=2\pi(4.6\pm0.3)$ MHz deduced from the knowledge of the input power into the cavity and the mirror transmissions. 

The good quantitative agreement shows that fits of the EIT spectra to the theoretical model can be used to extract reliable interaction rates. Note also that the clearly non-Lorentzian lineshape of the EIT transparency window, observed in Fig.~\ref{fig:rabi_EIT_spectra}b and which is a consequence of the inhomogeneous control field in our all-cavity EIT scheme, is perfectly reproduced by the theoretical model.

\begin{figure}
\centering
\includegraphics[width=0.5\textwidth]{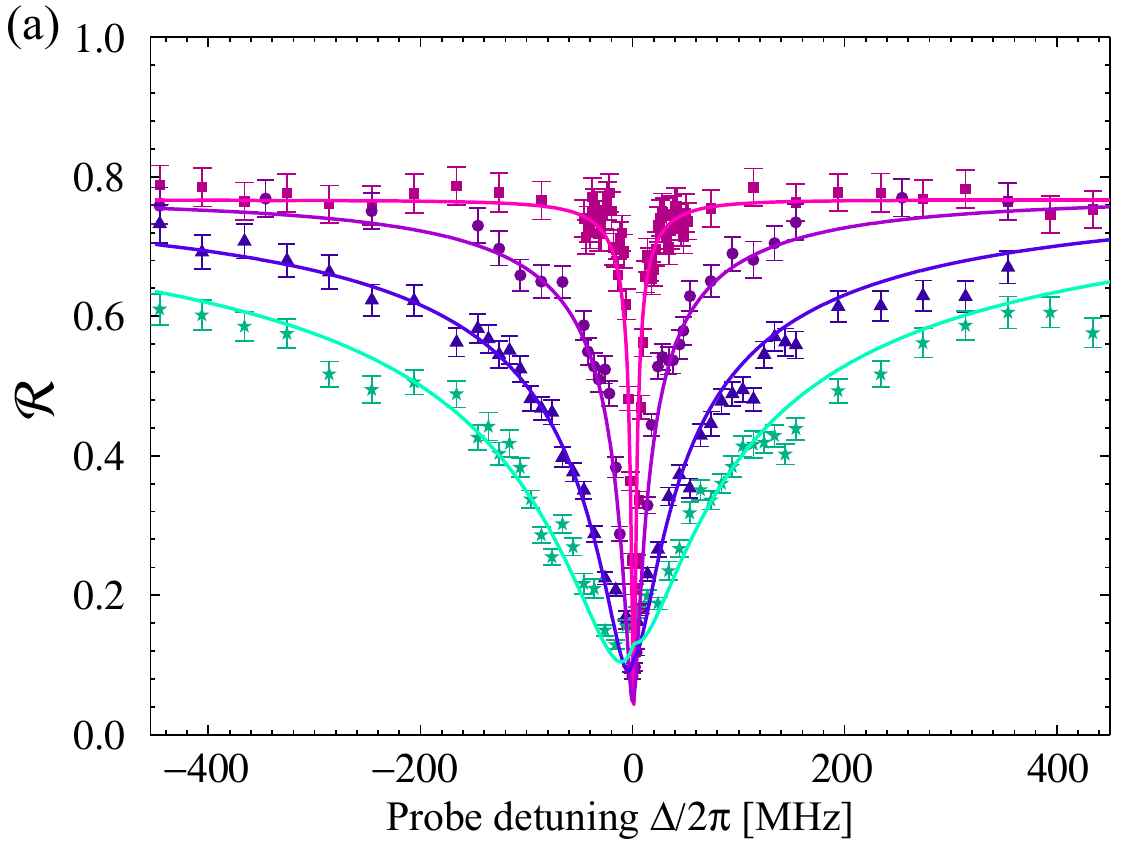}\\\vspace{5pt}
\includegraphics[width=0.5\textwidth]{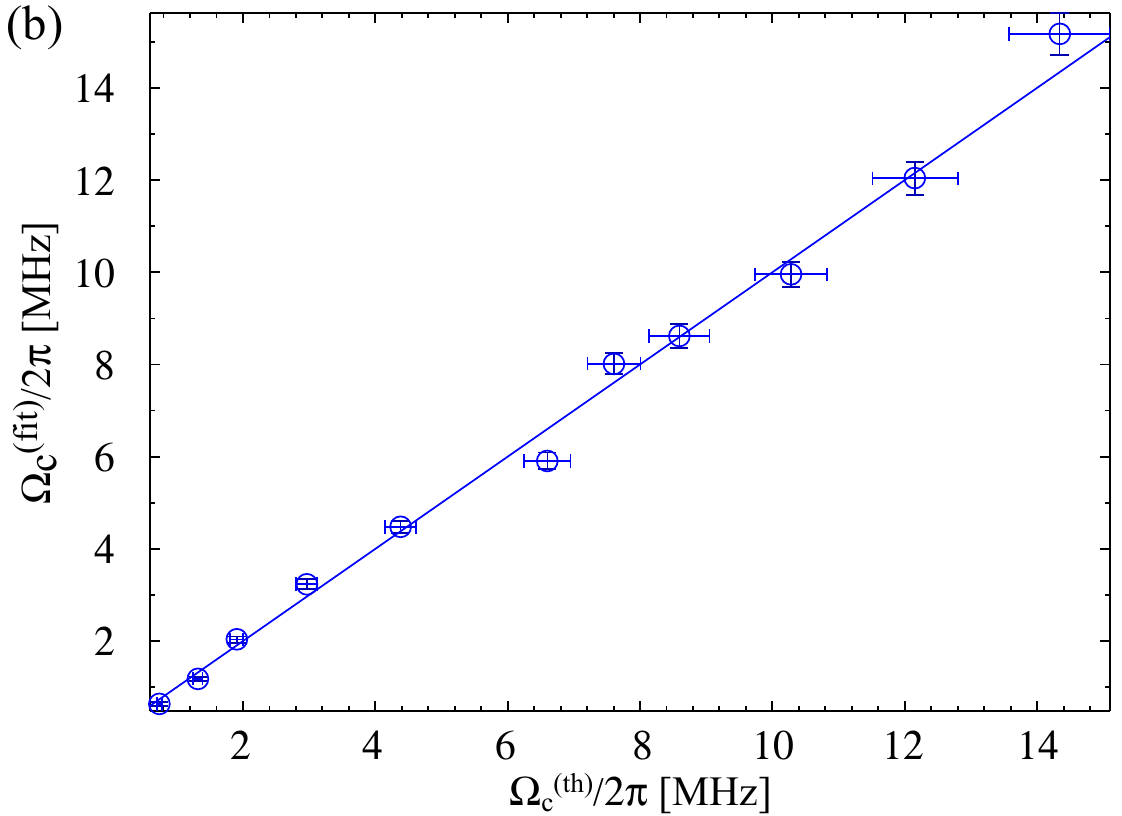}
\caption{(a) Probe reflectivity spectra for a cavity containing an ICC with cooperativity $C\simeq 5.4$ and for control fields with various Rabi frequencies. The lines show the results of theoretical fits, from which one obtains control field Rabi frequencies of $2\pi(1.18\pm0.04)$ MHz (red squares), $2\pi(3.23\pm0.11)$ MHz (purple dots), $2\pi(5.91\pm0.18)$ MHz (blue triangles) and $2\pi(8.62\pm0.26)$ MHz (green diamonds). (b) Control field Rabi frequencies as deduced from fits to the EIT spectra versus the Rabi frequencies expected from the injected laser power into the cavity. The solid line indicates $\Omega_c^{\textrm{(fit)}}=\Omega_c^{\textrm{(th)}}$.} \label{fig:EIT_dip_power}
\end{figure}

\subsection{Effect of control field intensity}

We now turn to more systematic investigations of the effect of the control field intensity on the EIT dynamics. First, steady state reflectivity spectra were measured for a slightly bigger crystal with an effective number of ions of $\sim890$, resulting in an expected collective coupling rate $g_N^{\textrm{(th)}}=2\pi(16.6\pm0.4)$ MHz and cooperativity $C\sim 5.4$. Some of the obtained transparency windows are shown in Fig.~\ref{fig:EIT_dip_power}a, together with the resulting fits to the theoretical models. A global fit to the data gives a collective coupling rate of $g_N=2\pi(16.2\pm 0.2)$ MHz and the control field Rabi frequencies $\Omega_c^{\textrm{(fit)}}$ shown in Fig.~\ref{fig:EIT_dip_power}b, which are in very good agreement with the expected ones, $\Omega_c^{\textrm{(th)}}$. Based on previous studies~\cite{Herskind2009NatPhys,Albert2011,Albert2012} a value of $\gamma_0=(2\pi) 1$ kHz for the ground state decoherence rate was used for the fit. Let us stress that, given the EIT timescales considered here being typically much shorter than the ground state decoherence time, the exact value of $\gamma_0$ is not critical for the analysis.

\begin{figure}
\centering
\includegraphics[width=0.5\textwidth]{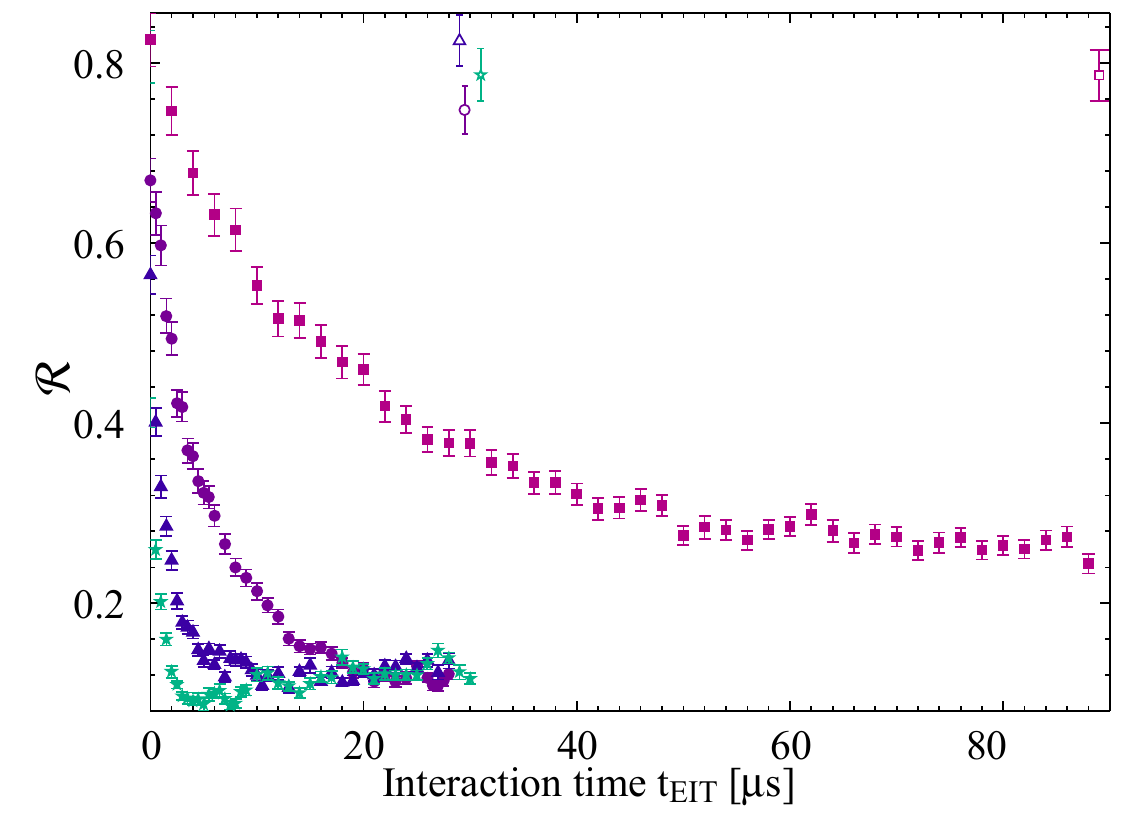}
\caption{Probe reflectivity on two-photon resonance as a function of the EIT interaction time $t_{\textrm{EIT}}$ for control fields with the same Rabi frequencies as in Fig.~\ref{fig:EIT_dip_power}a. The empty symbols show the probe reflectivity levels right after the control field has been switched off.} \label{fig:EIT_dynamics}
\end{figure}

While the previous data give a steady state picture of the effect of EIT interaction on the probe reflectivity spectrum, it is possible to dynamically observe the transient behavior of the buildup of the transparency by measuring the reflectivity level at different times during the EIT interaction. For this purpose we run an experimental sequence which is identical to the previous one, the only difference being that the APD probing time occuring after a time $t_{\textrm{EIT}}$ is reduced to $t_{\textrm{APD}}=0.5$ $\mu$s. In these experiments the probe field is kept on two-photon resonance. The resulting dynamics are shown in Fig.~\ref{fig:EIT_dynamics} for the same ICC and control field Rabi frequencies as in Fig.~\ref{fig:EIT_dip_power}. After application of the control field the probe reflectivity level is observed to decrease, all the more so that the control field is intense, then go through a minimum or not, depending on the atomic transparency level, and finally reach its steady state value. To check that the change in reflectivity and the buildup of a steady state transparency is effectively due to an EIT interaction, and not to a loss atoms from state $|1\rangle$, the control field is abruptly switched off at the end of the EIT interaction. The probe field reflectivity is then seen to return to its intial level (empty symbols in Fig.~\ref{fig:EIT_dynamics}), corresponding to maximal atomic absorption.

\begin{figure}
\centering
\subfigure{
\resizebox*{6.5cm}{!}{\includegraphics{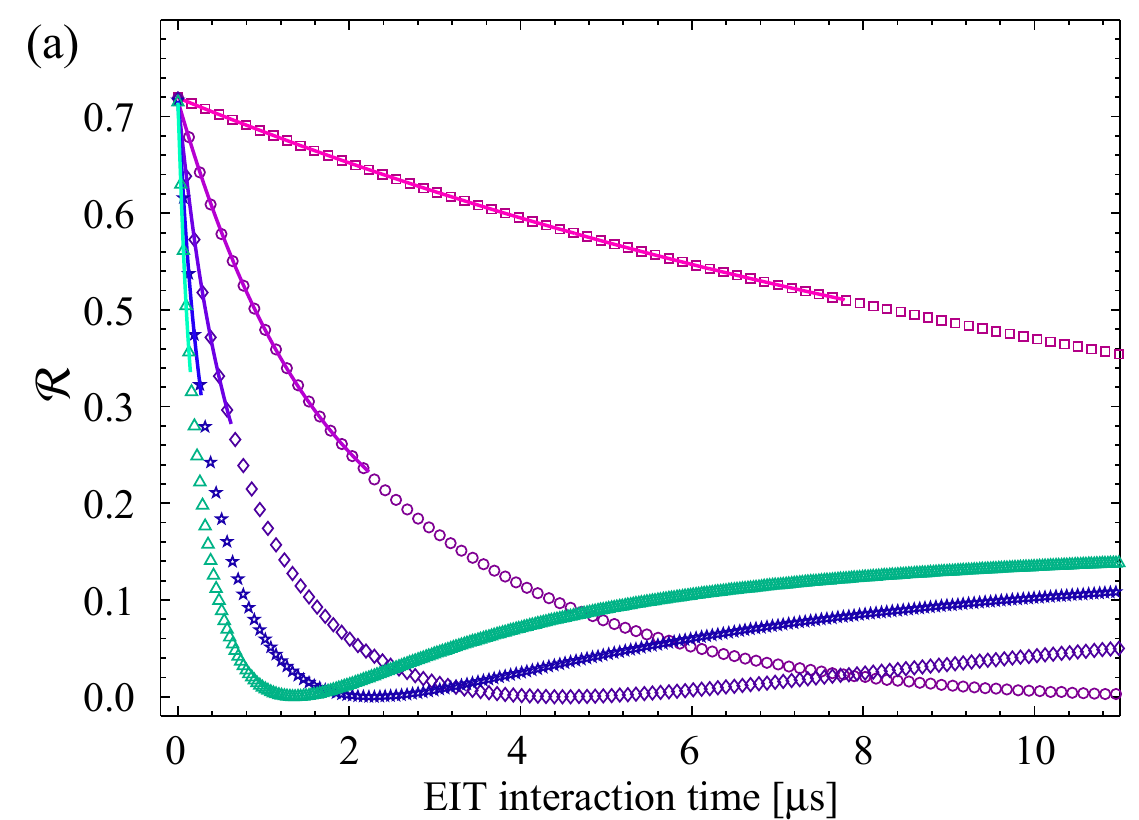}}}\hspace{5pt}
\subfigure{
\resizebox*{6.5cm}{!}{\includegraphics{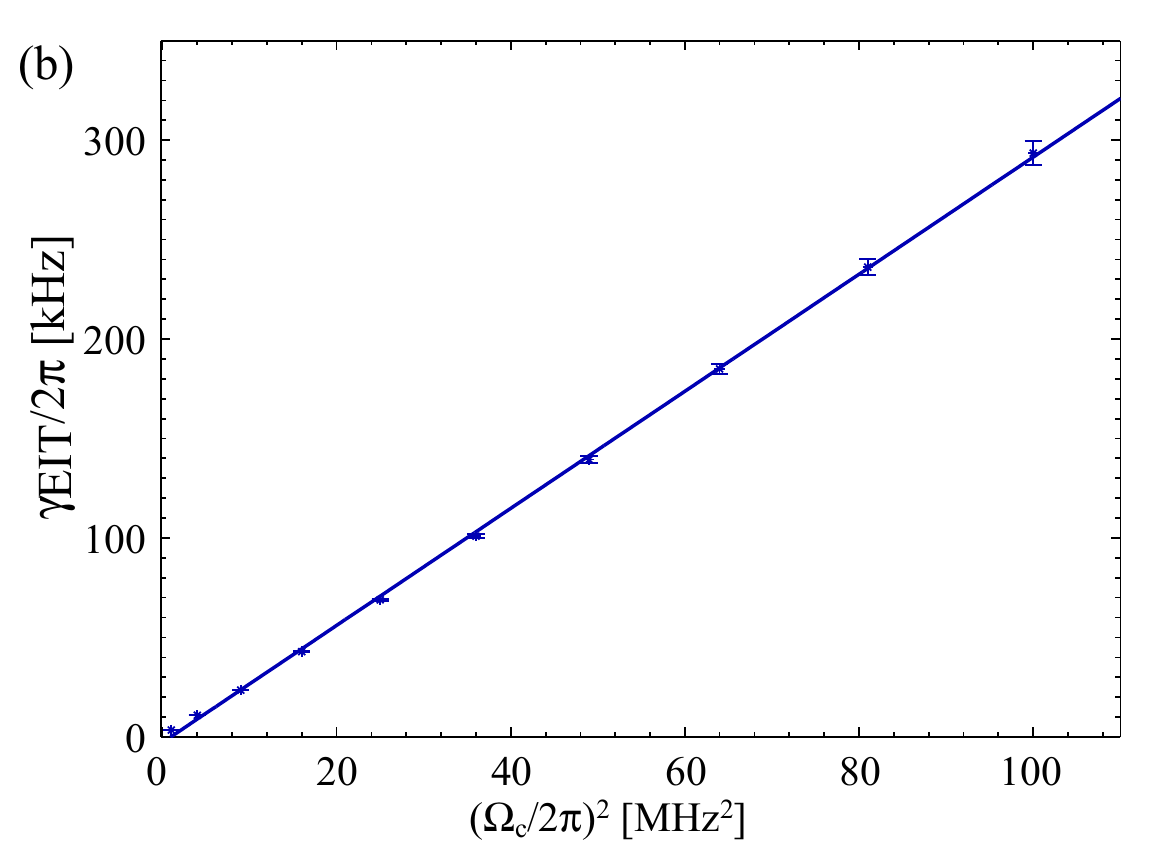}}}
\caption{(a) Simulated probe reflectivity dynamics in presence of a control field with various Rabi frequencies [$\Omega_c=(2\pi)1$ MHz (squares), $\Omega_c=(2\pi)3$ MHz (circles), $\Omega_c=(2\pi)5$ MHz (diamonds), $\Omega_c=(2\pi)7$ MHz (stars), $\Omega_c=(2\pi)9$ MHz (triangles)], for an ICC with cooperativity $C=5.4$ and on two-photon resonance. (b) Effective exponential decay rates $\gamma_{\textrm{EIT}}$ obtained from exponential fits of the simulated reflectivity curves for the first microseconds of the EIT interaction, as a function of the control field Rabi frequency squared. The solid lines in 8a) show the result of the exponential fits and in (b) the result of a linear fit.} \label{fig:EIT_dynamics_sim}
\end{figure}

As discussed in detail in~\cite{Dantan2012}, the dynamical evolution of the probe field in this ``all-cavity" EIT situation is complex due to the inhomogeneous profile of the control field in the interaction region, but can be numerically simulated. An example of such simulations is shown in Fig.~\ref{fig:EIT_dynamics_sim}. To facilitate the comparison between the predictions of the dynamical model and the experimental data the simulated data is fitted with an exponentially decaying function $\mathcal{R}=b\exp(-2\gamma_{\textrm{EIT}}t)$ for the first few microseconds of the EIT interaction. The resulting decay rate is observed in Fig.~\ref{fig:EIT_dynamics_sim}b to scale fairly linearly with the square of the control field Rabi frequency. The resonant EIT buildup rate can thus be approximated, at least in the beginning of the interaction, by $\gamma_{\textrm{EIT}}=\gamma_0+\frac{\Omega_c^2/\alpha}{2(1+2C)}$, where $\alpha$ is an effective scaling factor accounting for the control field Rabi frequency not being constant over the interaction area. In the range of Rabi frequencies considered, a fit of the simulated data yields $\alpha^{\textrm{(sim)}}=2.19\pm0.02$.

\begin{figure}
\centering
\includegraphics[width=0.5\textwidth]{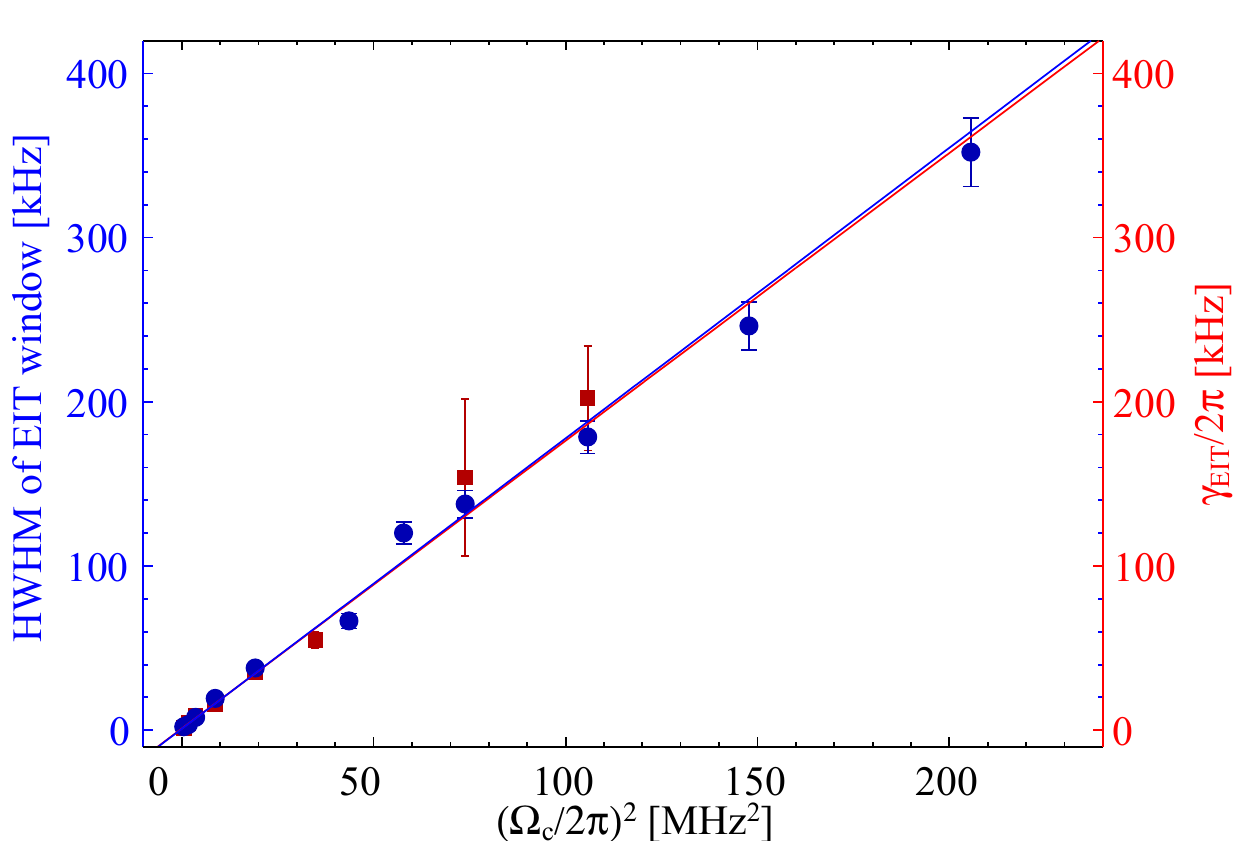}
\caption{HWHM of the EIT transparency windows (blue dots) shown in Fig.~\ref{fig:EIT_dip_power} and EIT buildup rates (red squares) obtained from short-time exponential fits of the curves shown in Fig.~\ref{fig:EIT_dynamics}, as a function of the control field Rabi frequency squared. The solid lines show the result of linear fits.} \label{fig:EIT_widths}
\end{figure}

Using the same approach to fit the experimental data of Fig.~\ref{fig:EIT_dynamics}a one extracts the EIT buildup rates shown as red squares in Fig.~\ref{fig:EIT_widths}. These rates are also observed to scale approximately linearly with the control field intensity, yielding an effective scaling factor $\alpha^{\textrm{(exp)}}=2.2\pm 0.2$. An excellent agreement is also observed between the scalings of the steady state HWHMs and the EIT buildup rates with the square of the control Rabi frequency, which confirms the complementarity between the time- and frequency-domain pictures.  Linear fits yield a slope of $(1.7\pm 0.1)\times 10^{-3}/(2\pi\textrm{MHz})$ and offset of $2\pi(0.9\pm0.3)$ kHz for the HWHMs, and a slope of $(1.8\pm 0.2)\times 10^{-3}/(2\pi\textrm{MHz})$ and offset of $2\pi(1.0\pm0.4)$ kHz for the EIT buildup rates. The offset is also consistent with the value of $\gamma_0$ chosen for the fits of the steady state spectra. Note also that this agreement is observed over a range of control field intensities giving rise to transparency windows with HWHMs ranging from $\sim  3$ to $\sim 350$ kHz, corresponding to a reduction of a factor up to $\sim650$ of the bare cavity linewidth. Let us finally point out that a good agreement between the frequency- and time-domain experimental data need not be expected {\it a priori}, since technical drifts and noises could affect the short- and long-term dynamics in different ways.

\subsection{Effect of the effective number of ions}

\begin{figure}
\centering
\subfigure{
\resizebox*{6.5cm}{!}{\includegraphics{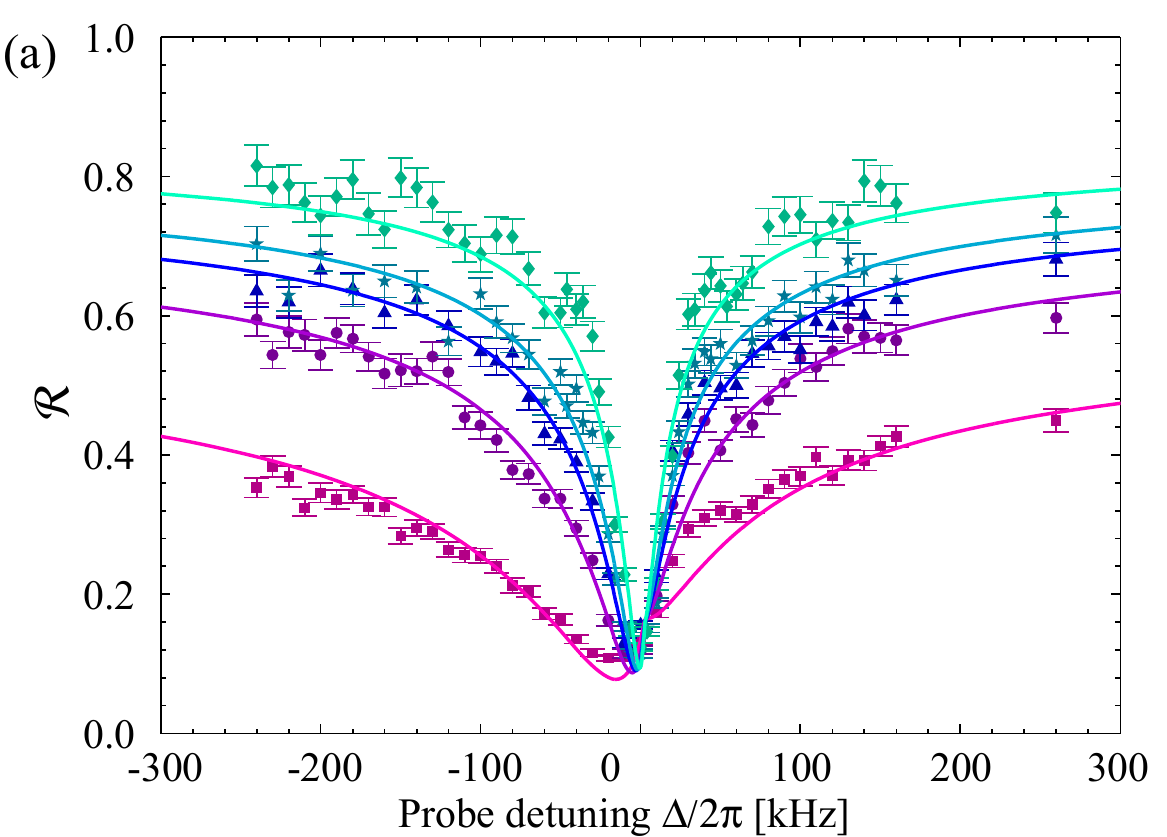}}}
\subfigure{
\resizebox*{5.7cm}{!}{\includegraphics{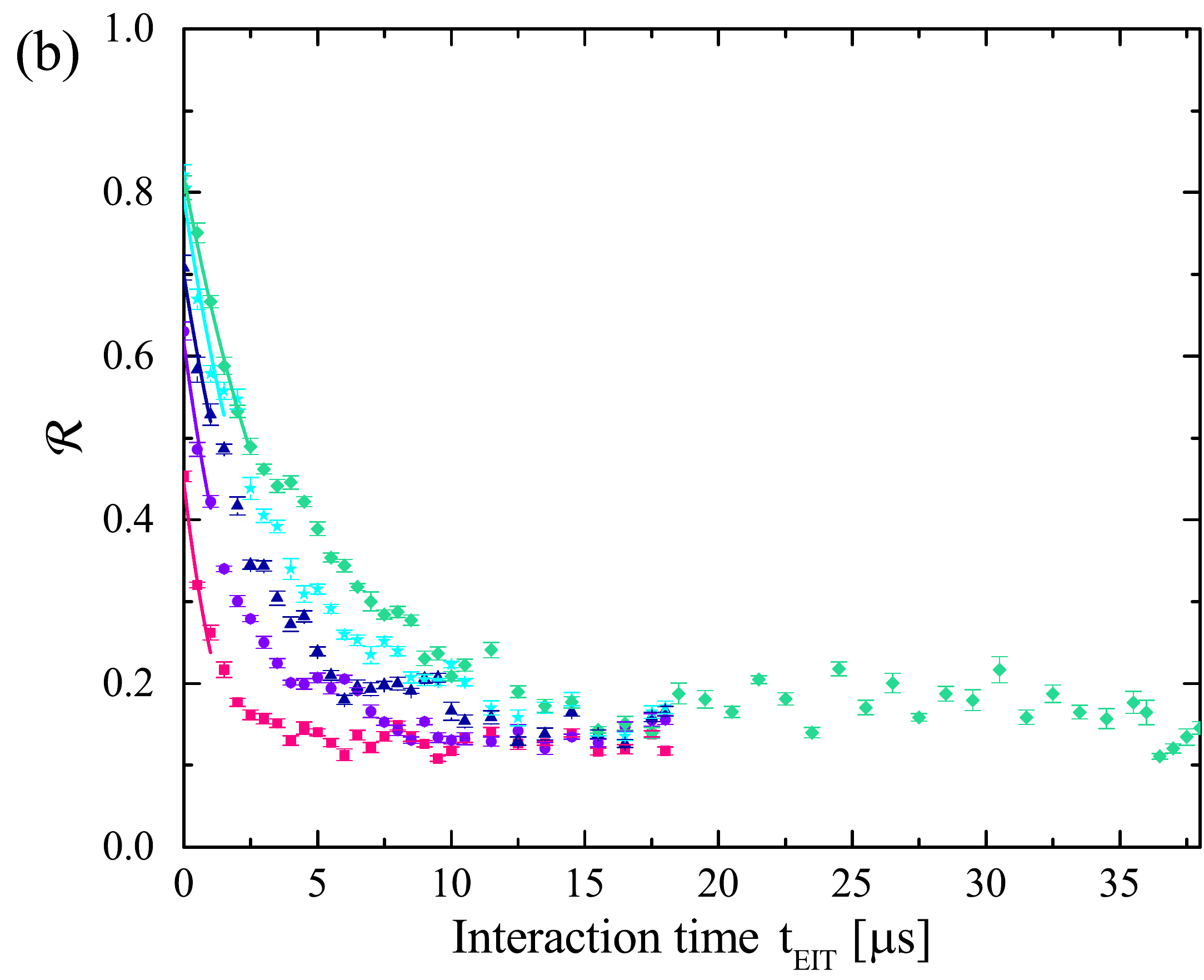}}}
\caption{(a) Probe reflectivity spectra for a cavity containing ICCs with different effective number of ions [$N=393$ (squares), $N=590$ (circles), $N=737$ (triangles), $N=938$ (stars), $N=1112$ (diamonds)] and for a control field with  Rabi frequency $\Omega_c=2\pi(3.5\pm0.2)$ MHz. The lines show the results of theoretical fits. (b) Probe reflectivity dynamics on two-photon resonance for a cavity containing ICCs with similar ion numbers [$N=366$ (squares), $N=575$ (circles), $N=738$ (triangles), $N=947$ (stars), $N=1047$ (diamonds) and the same control field Rabi frequency as in (a).} \label{fig:EIT_dipvsN}
\end{figure}

\begin{figure}
\centering
\includegraphics[width=0.5\textwidth]{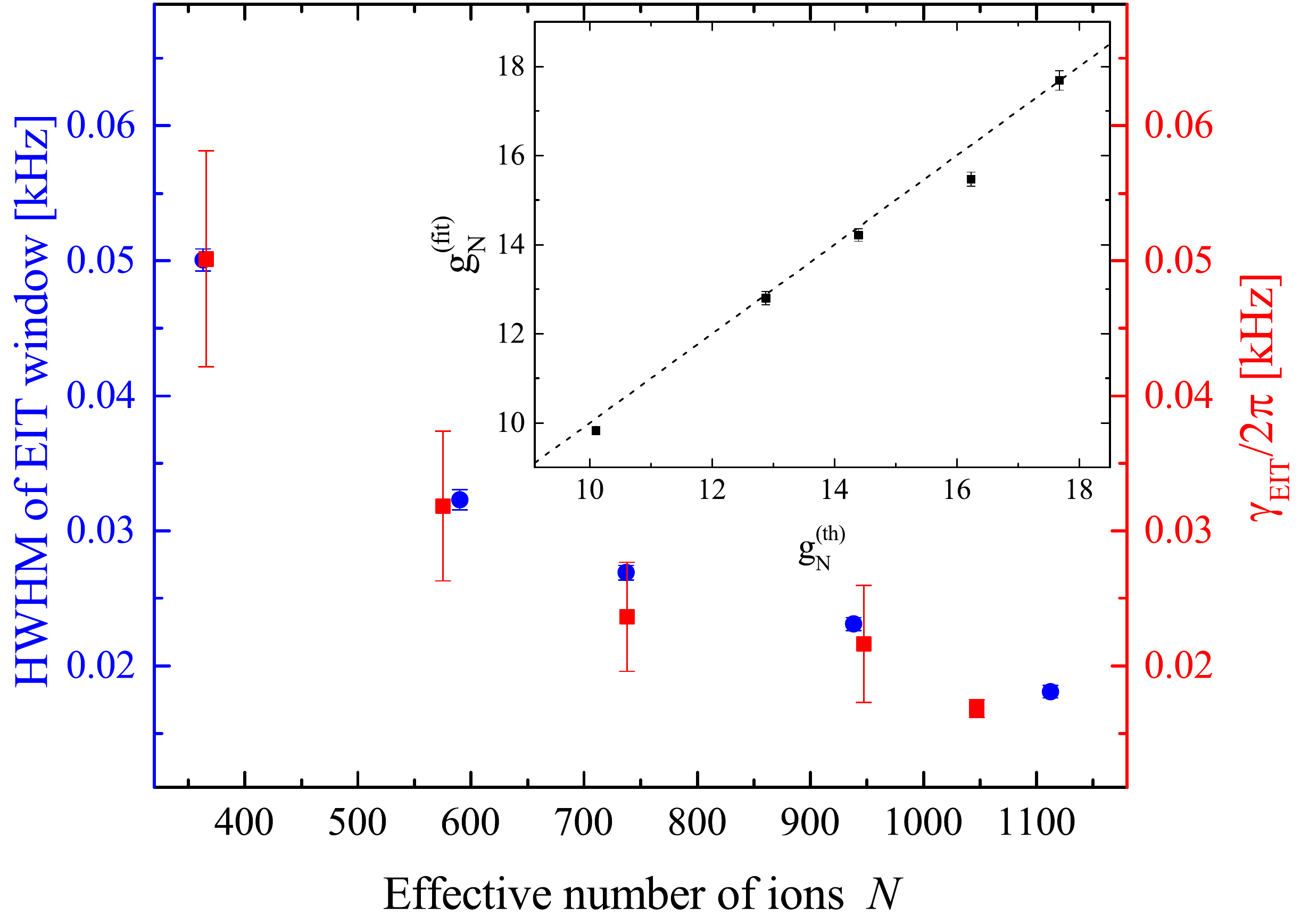}
\caption{HWHM of the EIT transparency windows (blue dots) and effective EIT buildup rate (red squares) as a function of the effective number of ions $N$ for the ICCs of Fig.~\ref{fig:EIT_dipvsN}. The inset shows the collective coupling rates deduced from fits of the spectra to the theoretical models versus the ones expected from the estimation of the number of ions. The solid line indicates $g_N^{\textrm{(fit)}}=g_N^{\textrm{(th)}}$.} \label{fig:EIT_widthsN}
\end{figure}

Complementary experiments were performed by varying the effective number of ions interacting with the fields for fixed values of the control Rabi frequency. The number of ions was varied by changing the axial confinement of the trap, while keeping the radiofrequency field amplitude, and thereby the density, constant. Figure~\ref{fig:EIT_dipvsN}a shows the resulting EIT transparency windows (blue dots) for crystals with effective number of ions ranging from $\sim360$ to $\sim1100$ and for a control field Rabi frequency of $\Omega_c=(2\pi)3.5$ MHz.  As expected, one observes a narrowing of the transparency window as the effective number of ions - \textit{i.e.}, the optical depth of the medium - is increased. Figure~\ref{fig:EIT_dipvsN}b shows the complementary time-picture of the EIT buildup on two-photon resonance for ICCs having approximately the same effective number of ions. Fitting the first few $\mu$s of evolution with an exponentially decaying function yields the effective EIT buildup rates (red squares) displayed in Fig.~\ref{fig:EIT_widthsN}. Again, good agreement is observed between the short-term dynamics rates and the widths of the steady state transparency windows.

From global fits of the spectra data to the theoretical model the collective coupling rates $g_N^{\textrm{(fit)}}$ are extracted for each crystal and compared with the ones expected from the fluorescence image determination of the ion number, $g_N^{\textrm{(th)}}$ (inset of Fig.~\ref{fig:EIT_widthsN}). Very good agreement is again observed between the experimental data and the theoretical expectations, highlighting the high-level of control displaying by ICCs for cavity QED.

\section{Conclusion}\label{sec:conclusion}

We have investigated the transient dynamics of a cavity field at the single photon level under conditions of electromagnetically induced transparency in large ion Coulomb crystals. The steady state cavity spectrum as its dynamics during the EIT buildup were studied as a function of the control field intensity and for ICCs with various number of ions. The very good overall agreement between the experimental observations and the theoretical predictions for this specific ``all-cavity" EIT configuration confirm the applicability of the model and represent a nice verification of the complementary frequency- and time-domain pictures of cavity EIT dynamics. The results also highlight the high level of control on the light-matter interaction parameters that can be achieved with ion Coulomb crystals in cavities and are promising, \textit{e.g.}, for the realization of ICC-based photon memories and counters~\cite{Zangenberg2012,Clausen2013}.

\section*{Funding}

We acknowledge support from the Carlsberg Foundation, Villumfonden, the Danish Natural Science Research Council through the European Science Foundation EuroQUAM 'Cavity Mediated Molecular Cooling' project and the Sapere Aude initiative, and the STREP project 'Physics of Ion Coulomb Crystals' under the European Commission FP7 programme.


\bibliographystyle{tfp}
\bibliography{cavityEITdynamicsbiblio.bib}


\end{document}